\documentclass[lettersize,journal]{IEEEtran}
\usepackage{amsmath,amsfonts}
\usepackage{algorithmic}
\usepackage{array}
\usepackage[caption=false,font=normalsize,labelfont=sf,textfont=sf]{subfig}
\usepackage{textcomp}
\usepackage{stfloats}
\usepackage{url}
\usepackage{verbatim}
\usepackage{graphicx}
\usepackage{balance}
\usepackage{lipsum}
\usepackage[hidelinks]{hyperref}

\usepackage{tikz}
\usepackage{pgfplots}
\usetikzlibrary{backgrounds}
\usetikzlibrary{positioning}
\usetikzlibrary{patterns}
\usetikzlibrary{calc}
\usetikzlibrary{fit}
\usetikzlibrary{arrows.meta}
\pgfplotsset{compat=1.18}
\usepackage{adjustbox}

\usepackage{booktabs}
\usepackage{tabularx}

\usepackage[nameinlink,capitalize]{cleveref}
\crefname{table}{Tab.}{Tabs.}

\usepackage{siunitx}
\DeclareSIUnit{\symbol}{sym}
\DeclareSIUnit{\snu}{SNU}
\DeclareSIUnit{\db}{dB}

\usepackage[abbreviations]{glossaries-extra}
\GlsXtrEnableEntryCounting{abbreviation,acronym}{1}
\setabbreviationstyle[acronym]{long-short}
\newacronym{acr:cv-qkd}{CV-QKD}{continuous-variable quantum key distribution}
\newacronym{acr:dv-qkd}{DV-QKD}{discrete-variable quantum key distribution}
\newacronym{acr:its}{ITS}{information-theoretic security}
\newacronym{acr:snr}{SNR}{signal-to-noise ratio}
\newacronym{acr:skr}{SKR}{secret key rate}
\newacronym{acr:b2b}{B2B}{back-to-back}
\newacronym{acr:voa}{VOA}{variable optical attenuator}
\newacronym{acr:fer}{FER}{frame error rate}
\newacronym{acr:ldpc}{LDPC}{low-density parity-check}
\newacronym{acr:bi-awgn}{BI-AWGN}{binary-input additive white Gaussian noise}
\newacronym{acr:met}{MET}{multi-edge type}
\newacronym{acr:ir}{IR}{information reconciliation}
\newacronym{acr:dsp}{DSP}{digital signal processing}
\newacronym{acr:rl-ldpc}{RL-LDPC}{Raptor-like low-density parity-check}
\newacronym{acr:dac}{DAC}{digital-to-analog converter}
\newacronym{acr:adc}{ADC}{analog-to-digital converter}
\newacronym{acr:pa}{PA}{privacy amplification}
\newacronym{acr:rsa}{RSA}{Rivest–Shamir–Adleman}
\newacronym{acr:pqc}{PQC}{post-quantum cryptography}
\newacronym{acr:qkd}{QKD}{quantum key distribution}
\newacronym{acr:pm}{PM}{power meter}
\newacronym{acr:ps}{PS}{probabilistic-shaped}
\newacronym{acr:awgn}{AWGN}{additive white Gaussian noise}
\newacronym{acr:qrng}{QRNG}{quantum random number generator}
\newacronym{acr:llr}{LLR}{log-likelihood ratio}
\newacronym{acr:fec}{FEC}{forward error correction}
\newacronym{acr:mdr}{MDR}{multidimensional reconciliation}
\newacronym{acr:bpsk}{BPSK}{binary phase-shift keying}
\newacronym{acr:crc}{CRC}{cyclic redundancy check}
\newacronym{acr:snu}{SNU}{shot noise unit}
\newacronym{acr:awg}{AWG}{arbitrary waveform generator}
\newacronym{acr:abc}{ABC}{automatic bias control}
\newacronym{acr:iqm}{IQM}{I/Q-modulator}
\newacronym{acr:dp-iqm}{DP IQM}{dual-polarization I/Q-modulator}
\newacronym{acr:ecl}{ECL}{external cavity laser}
\newacronym{acr:pspr}{PSPR}{pilot-to-signal power ratio}
\newacronym{acr:ssmf}{SSMF}{standard single mode fiber}
\newacronym{acr:llo}{LLO}{local-local oscillator}
\newacronym{acr:fm-qpsk}{FM-QPSK}{frequency-multiplexed QPSK}
\newacronym{acr:bpd}{BPD}{balanced photo diode}
\newacronym{acr:dso}{DSO}{digital sampling oscilloscope}
\newacronym{acr:lti}{LTI}{linear time-invariant}
\newacronym{acr:spa}{SPA}{sum-product algorithm}
\newacronym{acr:pt}{PT}{pilot-tone}
\newacronym{acr:psd}{PSD}{power spectral density}
\newacronym{acr:lo}{LO}{local oscillator}

\usepackage[backend=biber, style=ieee, maxnames=3, minnames=1]{biblatex}
\DefineBibliographyStrings{english}{
  url    = Available (commit: \texttt{edf3450b}),
}

\DeclareRobustCommand{\circled}[1]{%
  \leavevmode
  \raisebox{.5pt}{\textcircled{\raisebox{-.9pt}{#1}}}%
}

\begin{document}

\title{Practical Methods for Distance-Adaptive Continuous-Variable Quantum Key Distribution}
\author{
  Jonas~Berl,~\IEEEmembership{Graduate Student Member,~IEEE,}
  Utku~Akin,~\IEEEmembership{Graduate Student Member,~IEEE,}
  Erdem~Eray~Cil,~\IEEEmembership{Graduate Student Member,~IEEE,}
  Laurent~Schmalen,~\IEEEmembership{Fellow,~IEEE,} \\
  and Tobias~Fehenberger,~\IEEEmembership{Member,~IEEE}
  \thanks{%
    This work is partially funded by the German Federal Ministry of Research, Technology and Space under the project DE-QOR (Grant IDs 16KISQ052K and 16KISQ056) and SEQUIN (Grant ID 16KlS2132K).
    This article was presented in part at the European Conference on Optical Communication (ECOC), Frankfurt, Germany, September 2024~\cite{Berl2024}.
    (Corresponding author: Jonas Berl (\href{mailto:jonas.berl@advasecurity.com}{jonas.berl@advasecurity.com}))

    Jonas Berl, Utku Akin and Tobias Fehenberger are with Adva Network Security GmbH, Fraunhoferstraße 5, Martinsried/Munich, Germany.

    Jonas Berl, Erdem Eray Cil and Laurent Schmalen are with the Communications Engineering Lab,  Karlsruhe Institute of Technology, Hertzstraße 16, Karlsruhe, Germany.

    Utku Akin is with the Institute for Communications Engineering, Technical University of Munich, Theresienstraße 90, Munich, Germany.

    Article DOI: \href{https://doi.org/10.1109/JLT.2026.3664474}{10.1109/JLT.2026.3664474}

    © 2026 IEEE. Personal use of this material is permitted. Permission from IEEE must be obtained for all other uses, in any current or future media, including reprinting/republishing this material for advertising or promotional purposes, creating new collective works, for resale or redistribution to servers or lists, or reuse of any copyrighted component of this work in other works.
  }
}

\markboth{Journal of Lightwave Technology}{How to Use the IEEEtran \LaTeX \ Templates}

\maketitle

\begin{abstract}
    \Gls{acr:cv-qkd} is a promising quantum-safe alternative to classical asymmetric cryptography that enables two authenticated parties to establish a shared secret over a potentially eavesdropped quantum channel.
    A key step in \gls{acr:cv-qkd} post-processing is information reconciliation, which leverages \gls{acr:fec} techniques to extract identical bit strings from noisy correlated data.
    In this work, we analyze the strict limitations on operating distance that are imposed by constant-rate \gls{acr:fec}, severely limiting the practicability of \gls{acr:cv-qkd} systems in deployed optical networks.
    To overcome the distance limitations, we evaluate three strategies: (i) tuning modulation variance, (ii) adding controlled amounts of trusted detector loss, and (iii) the use of rate-adaptive \gls{acr:fec}.  
    All approaches are validated experimentally, compared in terms of performance, and we discuss implementation aspects.
    Our results show that while methods (i) and (ii) extend the operational distance of constant-rate \gls{acr:fec} without the need for additional hardware components, they incur a significant penalty in \gls{acr:skr}.
    In contrast, rate-adaptive \gls{acr:fec} enables \gls{acr:cv-qkd} operation with performance close to the asymptotic \gls{acr:skr} over a wide range of distances, provided that the reconciliation efficiency is chosen appropriately.
\end{abstract}

\begin{IEEEkeywords}
    CV-QKD, information reconciliation, distance-adaptivity, secret key rate
\end{IEEEkeywords}

\glsresetall

\section{Introduction}\label{sec:introduction}
\IEEEPARstart{I}{n} recent years, quantum computing has seen remarkable advances~\cite{Gyongyosi2019,Gill2022,Yang2023}, posing a significant threat to secure communication in the foreseeable future.
With Shor's algorithm and cryptographically-relevant quantum computing, asymmetric schemes such as \acrfull{acr:rsa}, the Diffie-Hellman-Merkle key agreement or elliptic curve cryptography are at high risk~\cite{Shor1997,Gidney2025}.
As a result, attention has shifted to quantum-safe cryptographic schemes, such as \gls{acr:pqc} and \gls{acr:qkd}.
While \gls{acr:pqc} is an algorithmic solution and is therefore still susceptible to future advances in mathematics and quantum computing, \gls{acr:qkd} offers the potential for \gls{acr:its}~\cite{Pirandola2020}.

Since its introduction by Bennett and Brassard in 1984~\cite{Bennett2014}, \gls{acr:qkd} has seen significant development with the first commercial systems available as of today.
\gls{acr:qkd} is a method for \textit{key distribution}, i.e., it is meant to establish a shared secret key between two authenticated parties, typically connected by an optical fiber or a free space optical link.
There are two main types of \gls{acr:qkd}.
In \gls{acr:dv-qkd}, the transmitting side, often called Alice, encodes random information on discrete variables of a quantum state, e.g., on the polarization of single photons, and sends the states to the receiving side, typically named Bob~\cite{Pirandola2020}.
The other type is \gls{acr:cv-qkd}, where Alice encodes the information on continuous variables of a coherent state, e.g., in the quadratures of an electromagnetic field emitted by a laser~\cite{Laudenbach2018}.
The security of both methods is based on quantum mechanical principles, such as the no-cloning theorem and the uncertainty principle~\cite{Scarani2009}.
These forbid a potential eavesdropper, Eve, from obtaining a (partial) copy of the transmitted information without being detected.
Due to its compatibility with commercial off-the-shelf components and photonic miniaturization~\cite{Zhang2019,Aldama2022,Hajomer2024}, \gls{acr:cv-qkd} is regarded as a cost-effective alternative to \gls{acr:dv-qkd}.
However, \gls{acr:cv-qkd} comes with more challenges in post-processing, with the error correction step, commonly referred to as \gls{acr:ir}, being particularly more demanding.

In experimental investigations of \gls{acr:cv-qkd}, the emphasis is typically placed on physical layer effects, whereas the influence of post-processing on the \gls{acr:skr} is frequently not taken into account.
For example, the reconciliation efficiency and \gls{acr:fer} are often treated as constant values and little justification is provided for the specific choices made~\cite{Almeida2021,Roumestan2021,Pan2022,Pereira2022}.
In practice, however, both parameters change with the channel loss, are susceptible to temporal variations of channel conditions, and depend on the characteristics of the employed \gls{acr:fec}.
Besides an inaccurate estimation of the \gls{acr:skr}, the effects of distance-dependent \gls{acr:fer} and reconciliation efficiency also cause strict distance limitations.
For operators that expect \gls{acr:qkd} systems to function reliably over a wide range of distances, distance-adaptive \gls{acr:cv-qkd} solutions are therefore of particular interest.

Recent research has examined the influence of practical reconciliation on system performance, especially addressing the resulting distance limitations.
For instance, it has been demonstrated that an optimization of the modulation variance can support \gls{acr:cv-qkd} over a broad range of distances, even when only a limited number of constant-rate \gls{acr:fec} codes are available~\cite{Ma2023,Almeida2023a,Almeida2023b}.
Similarly, the controlled injection of trusted detector noise has been investigated numerically to match the \gls{acr:fec} code rate and received \gls{acr:snr}, outperforming alternative techniques for rate-adaption such as shortening and puncturing~\cite{Kreinberg2020}. 
As another option, rate-adaptive \gls{acr:fec} techniques, e.g., based on Raptor codes and \gls{acr:rl-ldpc} codes, have been proposed.
For example, simulations with Raptor codes show that joint optimization of modulation variance and rate-adaptive \gls{acr:fec} can yield high reconciliation efficiencies and positive \glspl{acr:skr} over a wide range of transmission distances~\cite{Zhou2019}.
To address small-scale channel fluctuations, the combination of controlled amounts of trusted detector noise and a rate-adaptive \gls{acr:fec} code has been shown effective for codes with step-wise rate adaption~\cite{Yang2024}.
Nonetheless, most of these approaches have not yet been validated by experiments, and a comprehensive comparison of these methods is missing.

In this paper, we illustrate the impact of practical reconciliation on \gls{acr:cv-qkd} system performance, focusing on the imposed distance limitations.
We validate our findings experimentally and investigate three practical strategies to achieve distance-adaptive operation.
The first method relies on tuning the modulation variance, so that the received \gls{acr:snr} is matched to the rate of the available \gls{acr:fec} code, enabling a single constant-rate \gls{acr:fec} code to be used across a broad range of distances.
The second approach leverages the trusted-device scenario, where deliberately adding a controlled amount of trusted detector loss can be used to similarly adjust the received \gls{acr:snr} for distance-adaptive operation.
As third strategy, we consider rate-adaptive \gls{acr:fec}, where the code rate used during \gls{acr:ir} can be adapted to the transmission distance.
We demonstrate all investigated methods experimentally, compare their performance, and discuss implementation aspects.

\section{Distance-Adaptive Reconciliation}\label{sec:distance-adaptivity}
\begin{figure*}
    \centering
    \begin{minipage}[t]{0.48\textwidth}
        \captionsetup{width=.5\linewidth}
        \centering
        \includegraphics{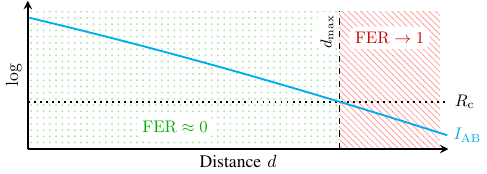}
        \caption{%
            Error correction for classical communications: mutual information $I_{\mathrm{AB}}$ and constant code rate $R_{\mathrm{c}}$ over the distance $d$ for a constant modulation variance.
        }
        \label{fig:ir_comms}
    \end{minipage}%
    \hfill
    \begin{minipage}[t]{0.48\textwidth}
        \centering
        \includegraphics{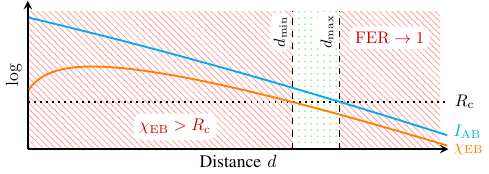}
        \caption{%
            \gls{acr:ir} for \gls{acr:cv-qkd}: mutual information $I_{\mathrm{AB}}$, Holevo bound $\chi_{\mathrm{EB}}$ and constant code rate $R_{\mathrm{c}}$ over the distance $d$ for a constant modulation variance.
        }
        \label{fig:ir_qkd}
    \end{minipage}
\end{figure*}

The main performance metric of any \gls{acr:qkd} system is the \gls{acr:skr} as a function of distance.
Considering reverse reconciliation, the asymptotic \gls{acr:skr} is computed as~\cite{Laudenbach2018}
\begin{equation}\label{eq:secret_fraction_asym}
    \mathrm{SKR} = \beta I_{\mathrm{AB}} - \chi_{\mathrm{EB}} \text{,}
\end{equation}
where $I_{\mathrm{AB}}$ is the mutual information between Alice and Bob, $\chi_{\mathrm{EB}}$ is the Holevo bound between Eve and Bob, and $\beta$ denotes the reconciliation efficiency.
When \gls{acr:ir} is implemented with \gls{acr:fec}, the reconciliation efficiency needs to be replaced by the multiplicative gap between code rate $R_{\mathrm{c}}$ and mutual information, i.e., \mbox{$\beta = R_{\mathrm{c}} / I_{\mathrm{AB}}$}.
We denote this updated key rate as \textit{extracted} \gls{acr:skr}.
Accounting for erroneous frames, the asymptotic extracted \gls{acr:skr} is formulated as~\cite{Laudenbach2018}
\begin{equation}\label{eq:secret_fraction_practical}
    \mathrm{SKR} = (1 - \mathrm{FER})(R_{\mathrm{c}} - \chi_{\mathrm{EB}}) \text{.}
\end{equation}
Here, we neglect other scaling factors, such as the fraction of quantum states used for parameter estimation or idle times during calibration.
\Cref{eq:secret_fraction_practical} helps explain the fundamental difference between error correction in classical communications and the impact of practical \gls{acr:ir} on the extracted \gls{acr:skr}.

According to the channel coding theorem, there exist error-correcting codes with decoder that allow an arbitrarily small error probability as long as the code rate $R_{\mathrm{c}}$ is smaller than the channel capacity (or equivalently the applicable mutual information)~\cite{Shannon1948}.
To illustrate the effect of constant-rate \gls{acr:fec}, we consider the mutual information $I_{\mathrm{AB}}$ of the symbols transmitted over the quantum channel between Alice and Bob.
For heterodyne detection, it is computed as~\cite{Laudenbach2018}
\begin{equation}\label{eq:mutual_information}
    I_{\mathrm{AB}} = \log_2\left(1 + \frac{T V_{\mathrm{mod}}}{2+\xi} \right)\text{,}
\end{equation}
where $T$ is the effective transmittance and includes both the channel and detector transmittance, $T_{\mathrm{ch}}$ and $T_{\mathrm{det}}$, respectively.
The modulation variance is denoted as $V_{\mathrm{mod}}$ and $\xi$ is the excess noise, both of which are typically normalized to the \acrfull{acr:snu}.

In \cref{fig:ir_comms}, we present a qualitative curve of mutual information as a function of the transmission distance, assuming a constant modulation variance and an infinitely steep \gls{acr:fer} degradation at channel capacity (i.e., we model the \gls{acr:fer} as a Heaviside step function).
The arbitrary but fixed code rate $R_{\mathrm{c}}$ of the \gls{acr:fec} code used during \gls{acr:ir} is depicted by the dotted horizontal line.
The channel attenuation is assumed to be \mbox{$T_{\mathrm{ch}} = 10^{-\alpha d}$}, where $\alpha$ is the fiber attenuation in \si{dB/km}.
As shown, as long as the mutual information is larger than the fixed code rate, codewords can be decoded successfully and the \gls{acr:fer} is approximately zero.
When the code rate becomes larger than the mutual information, codewords can no longer be decoded, resulting in an \gls{acr:fer} that approaches one. 
Therefore, the error correction step introduces an upper bound on the operating distance marked by the vertical dashed line at $d_{\mathrm{max}}$.
Beyond this distance, the correlated noisy data of Alice and Bob can no longer be reconciled.

For \gls{acr:cv-qkd} as stated in \eqref{eq:secret_fraction_practical}, two conditions must be fulfilled for a positive extracted \gls{acr:skr}.
First, at least one \gls{acr:fec} frame must be decoded successfully, which implies an \gls{acr:fer} of less than one.
Similarly to the upper bound on operating distance for classical communications, the error correction step limits the reach of the system.
Second and unique to \gls{acr:cv-qkd}, the rate of the employed \gls{acr:fec} code must be larger than the Holevo bound.
Otherwise, Eve obtains more information on the raw key material than Alice and Bob, and the protocol must be aborted.
Due to the course of the Holevo bound shown in \cref{fig:ir_qkd}, this condition is not fulfilled up to a minimum distance $d_{\mathrm{min}}$, marked by the left vertical dashed line.
Hence, only a small window of consecutive distances of width \mbox{$\Delta d = d_{\mathrm{max}} - d_{\mathrm{min}}$} can be covered with a single constant-rate \gls{acr:fec} code.
Moreover, as the difference \mbox{$I_{\mathrm{AB}} - \chi_{\mathrm{EB}}$} becomes smaller over distance, \gls{acr:fec} codes with increasingly higher reconciliation efficiencies are required to extract secrecy in long-distance scenarios.

The distance limitations of constant-rate \gls{acr:fec} pose severe practical challenges during the integration of \gls{acr:cv-qkd} into existing telecommunication infrastructure as operators expect systems to function reliably over a wide range of distances.
Trivially, a potentially large set of \gls{acr:fec} codes, i.e., one for each targeted distance window, can be used to overcome the distance limitations.
We refer to this strategy as method~\circled{0} and examine three alternative strategies in the following, summarized in \cref{tab:investigated_approaches}.

\begin{figure*}[t]
    \centering
    \begin{minipage}[t]{0.48\textwidth}
        \captionsetup{width=.5\linewidth}
        \centering
        \includegraphics{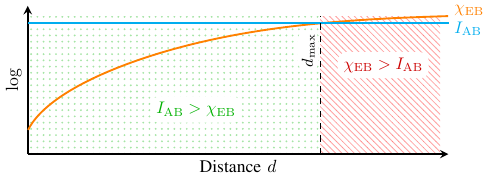}
        \caption{%
            Mutual information $I_{\mathrm{AB}}$ and Holevo bound $\chi_{\mathrm{EB}}$ over the distance $d$ when the modulation variance is tuned for constant received \gls{acr:snr}.
        }
        \label{fig:ir_vmod_tuning}
    \end{minipage}%
    \hfill
    \begin{minipage}[t]{0.48\textwidth}
        \centering
        \includegraphics{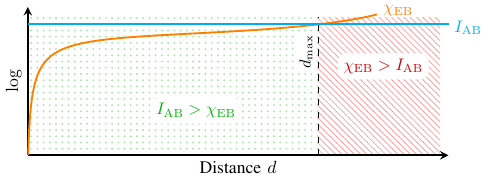}
        \caption{%
            Mutual information $I_{\mathrm{AB}}$ and Holevo bound $\chi_{\mathrm{EB}}$ over the distance $d$ when the trusted detector loss is scaled for constant received \gls{acr:snr}.
        }
        \label{fig:ir_tdet_tuning}
    \end{minipage}
\end{figure*}

In method~\circled{1}, the modulation variance is tuned for distance-adaptive operation~\cite{Almeida2023a,Ma2023}.
For this approach, we assume access to a single \gls{acr:fec} code with fixed rate $R_{\mathrm{c}}$.
Adjusting the modulation variance scales the signal power and thereby influences both the received \gls{acr:snr} and the Holevo bound.
By maintaining a constant \gls{acr:snr} at Bob's receiver, independent of channel transmittance, the \gls{acr:fer} condition in \eqref{eq:secret_fraction_practical} can be satisfied.
In \cref{fig:ir_vmod_tuning}, we depict the mutual information (or equivalently code rate) and Holevo bound qualitatively, scaling the modulation variance for a constant received \gls{acr:snr}.
As shown, there no longer is a lower limit on operational distance and the mutual information remains larger than the Holevo bound up to the maximum reach of the system marked by the dashed vertical line at $d_{\mathrm{max}}$.

The second strategy \circled{2} leverages the trusted-device scenario, in which the trusted parts of transmittance and excess noise, e.g., the inherent detector loss, are not attributed to an attacker during the computation of the \gls{acr:skr}~\cite{Laudenbach2018}.
In this case, the receiver can deliberately add trusted noise, trusted loss or a combination of both to match the received \gls{acr:snr} to the rate of the available \gls{acr:fec} code~\cite{Laudenbach2019,Kreinberg2020}.
Similarly to method~\circled{1}, we assume access to a single \gls{acr:fec} code with rate $R_{\mathrm{c}}$ and focus on trusted loss as this can be implemented with a \gls{acr:voa} at Bob's input.
Again, changing the trusted detector loss has an impact on the received \gls{acr:snr} and the Holevo bound.
It should be noted that although both tuning the modulation variance and adding trusted loss have the same effect on the received \gls{acr:snr}, their impact on the Holevo bound and consequently the \gls{acr:skr} differs, as illustrated in \cref{fig:ir_tdet_tuning}.
This difference arises because the Holevo bound is computed as the difference between two entropies: the von Neumann entropy of the state accessible to Eve and the von Neumann entropy of the same state after Bob has performed his measurement~\cite{Laudenbach2018}.
While tuning the modulation variance affects both entropies, adjusting the trusted detector loss only has an effect on the latter.

\begin{table}
    \centering
    \caption{Summary of investigated methods for distance-adaptive \gls{acr:cv-qkd}}
    \label{tab:investigated_approaches}
    \setlength{\tabcolsep}{10pt}
    \begin{tabularx}{0.95\columnwidth}{@{}lX@{}}
        \toprule
        Method & Description \\\midrule
        Method \circled{0} (trivial) & Set of constant-rate \gls{acr:fec} codes \\
        Method \circled{1} & Single constant-rate \gls{acr:fec} code with tuned modulation variance \\
        Method \circled{2} & Single constant-rate \gls{acr:fec} code with adding controlled amounts of trusted detector loss \\
        Method \circled{3} & Rate-adaptive \gls{acr:fec} code \\\bottomrule
    \end{tabularx}
\end{table}

As a third approach \circled{3}, we investigate the use of rate-adaptive \gls{acr:fec}, with an emphasis on \gls{acr:rl-ldpc} codes that have recently attracted interest in \gls{acr:cv-qkd}~\cite{Wang2018,Zhou2021}.
\gls{acr:rl-ldpc} codes are designed by first constructing a parity-check matrix for a high-rate code and then extending it to lower rates through the incremental addition of degree-\num{1} parity bits with optimized connectivity.
Compared to conventional methods for rate-adaptivity, such as puncturing or shortening, this approach yields superior performance across the low code rate regime typically considered for \gls{acr:cv-qkd}~\cite{Cil24}.
During encoding, the number of added degree-\num{1} parity bits is chosen according to the desired code rate.
These bits are generated based on pre-optimized connections and are appended to the high rate code word, thereby reducing the effective rate.
Decoding is done on a specific part of the graph representation of the code matching the lowest code rate, while the selection of the subgraph is determined by the target code rate.
This approach allows for a single decoder design capable of supporting multiple code rates, eliminating the need for distinct hardware implementations.

\begin{figure*}[t]
    \centering
    \adjustbox{max width=\textwidth}{
        \includegraphics{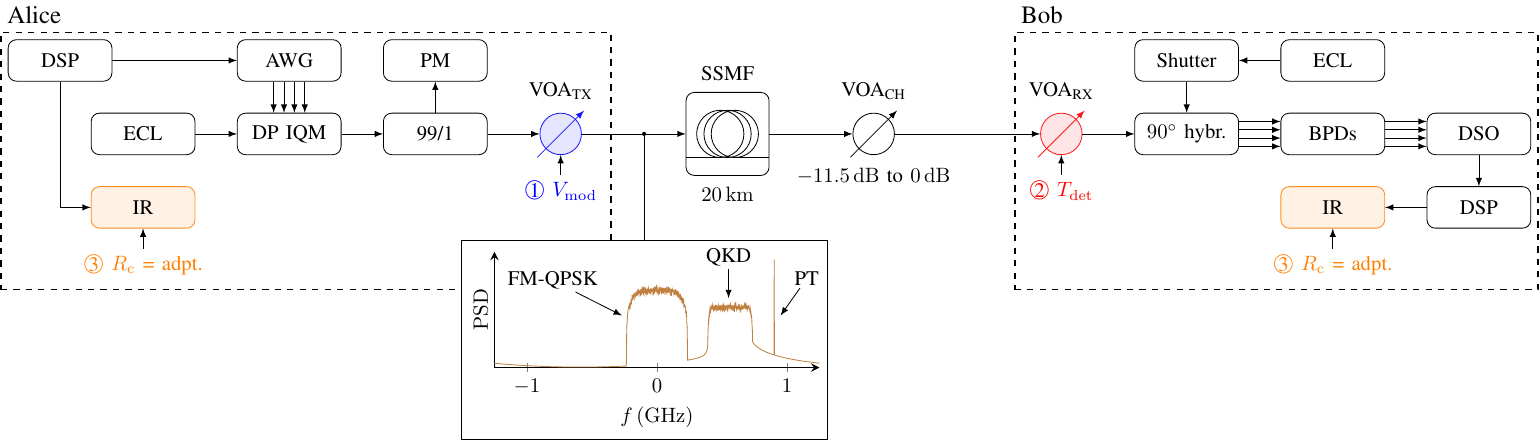}
    }
    \caption{%
        Experimental setup for distance-adaptive \gls{acr:cv-qkd}.
        The investigated approaches \protect\circled{1} to \protect\circled{3} are colored in blue, red and orange.
        The inset shows a qualitative plot of the \acrfull{acr:psd} of the transmit signal, consisting of \acrshort{acr:fm-qpsk}, \acrshort{acr:qkd} and the digital \acrshort{acr:pt}.
    }
    \label{fig:dist_adpt_cvqkd}
\end{figure*}

\section{Experimental Setup}\label{sec:experimental-setup}
The \gls{acr:cv-qkd} system under study is depicted in \cref{fig:dist_adpt_cvqkd}.
The transmitter consists of an offline \gls{acr:dsp} block which generates an eight-fold oversampled \gls{acr:ps}-256QAM transmit sequence with a symbol rate of \SI{312.5}{MBd} and a block length of \si{2^{19}} quantum states.
Experimentally, the use of higher-order \gls{acr:ps}-QAM has been shown to sufficiently approximate Gaussian modulation and achieve high \glspl{acr:skr} while staying compatible with practical equipment~\cite{Roumestan2024}.
To facilitate clock recovery and I/Q skew compensation on Bob's side, a \gls{acr:fm-qpsk} signal is added, while polarization de-multiplexing and carrier recovery are facilitated by an additional digital \gls{acr:pt}~\cite{Kleis2017,Laudenbach2019a,Chin2022}.
Both signals are known by the receiver and their power level relative to the \gls{acr:qkd} signal is defined as \gls{acr:pspr}, i.e.,
\begin{equation}
    \mathrm{PSPR} = \frac{P_{\mathrm{FM-QPSK}} + P_{\mathrm{PT}}}{P_{\mathrm{QKD}}} \text{,}
\end{equation}
where $P_{\mathrm{FM-QPSK}}$ denotes the signal power of the \gls{acr:fm-qpsk} signal, $P_{\mathrm{PT}}$ is the power of the digital pilot and $P_{\mathrm{QKD}}$ is the power of the QKD signal.
We use an \acrfull{acr:ecl} tuned to \SI{1550}{nm} with a spectral line width of less than \SI{2}{kHz} and a constant optical output power of \SI{13}{dBm}.
The transmit sequence consisting of \gls{acr:qkd}, \gls{acr:fm-qpsk} and digital pilot is generated with an \gls{acr:awg} at a sampling rate of \SI{2.5}{GHz}, while the electro-optical conversion happens in a \acrfull{acr:dp-iqm}.
To monitor the power of the \gls{acr:qkd} signal continuously, we utilize a 99/1 splitter and measure \SI{1}{\percent} of the incident power with an optical \acrfull{acr:pm}.
Since the transmit signal is generated in the high-power regime, we use the measured optical power and \gls{acr:voa}\textsubscript{TX} to ensure a constant modulation variance of \SI{5}{SNU} throughout all experimental protocol runs.

The quantum channel between Alice and Bob consists of \SI{20}{km} \gls{acr:ssmf} and \acrshort{acr:voa}\textsubscript{CH}.
Given the low signal powers employed in the experimental setup, effects from fiber non-linearity are negligible.
In addition, linear transmission impairments, including chromatic dispersion and polarization-mode dispersion, have no significant impact on the signal at the low symbol rate, and are therefore not mitigated in \gls{acr:dsp}.
Due to the stabilized experimental environment, state-of-polarization rotations remain largely static and are effectively compensated in \gls{acr:dsp}.
The attenuation of VOA\textsubscript{CH} can therefore be mapped to an equivalent \gls{acr:ssmf} distance, illustrating operation beyond \SI{20}{km}.

\begin{table}
    \centering
    \caption{Parameters of the experimental setup}
    \label{tab:experimental_params}
    \setlength{\tabcolsep}{13pt}
    \begin{tabularx}{0.7\columnwidth}{@{}ll@{}}
        \toprule
        Parameter & Value \\\midrule
        Symbol rate & \SI{312.5}{MBd} \\
        Block length & \si{2^{19}} \\
        Modulation format & PS-256 QAM \\
        Laser wavelength & \SI{1550}{nm} \\
        Laser linewidth & $\leq \SI{2}{kHz}$ \\
        Laser power & \SI{13}{dBm} \\
        Modulation variance & \SI{5}{SNU} \\
        \gls{acr:voa}\textsubscript{CH} attenuation & \SIrange{-11.5}{0}{dB} \\
        \gls{acr:dso} sampling rate & \SI{2.5}{GHz} \\
        \gls{acr:awg} sampling rate & \SI{2.5}{GHz} \\
        \gls{acr:fer} threshold & \num{1e-2} \\\bottomrule
    \end{tabularx}
\end{table}

On the receiving side, the signal enters another \gls{acr:voa} that is used in method~\circled{2} to control the trusted detector loss, while additionally functioning as a shutter during the calibration procedure.
We denote this \gls{acr:voa} as \acrshort{acr:voa}\textsubscript{RX} in the following and configure \SI{0}{\dB} attenuation during the experiments.
After passing through the optical \SI{90}{\degree} hybrid and the four \acrfullpl{acr:bpd}, the signal is captured with a \acrfull{acr:dso} at a sampling rate of \SI{2.5}{GHz}.
The laser used for the \gls{acr:llo}\footnote{Early \gls{acr:cv-qkd} systems transmitted the \gls{acr:lo} along the quantum signal from Alice to Bob. To emphasize that the \gls{acr:lo} is generated on Bob's side, the term \textit{local-local} oscillator is used.} on Bob's side is tuned to \SI{1550}{nm} and paired with Alice's transmit laser to minimize the carrier frequency offset between both lasers.
Additionally, the \gls{acr:llo} can be cut-off using a polarization-maintaining shutter during calibration.
The sampled signal is processed by offline \gls{acr:dsp}, making use of the \gls{acr:fm-qpsk} signal and digital \gls{acr:pt} embedded in the signaling scheme.
We calibrate the \gls{acr:qkd} system before each quantum transmission to minimize the effect of systematic deviations between calibration and quantum transmission as described in~\cite{Laudenbach2018}.
Both parties share the block for \gls{acr:ir}.
We follow the common assumption that Alice's and Bob's devices are trusted, i.e., an attacker cannot influence the device internals, marked by the dashed boundaries in \cref{fig:dist_adpt_cvqkd}.
All experimental parameters are summarized in \cref{tab:experimental_params}.

\subsection{Emulating SSMF Distance}
To study the effects of varying operating distances, we sweep the channel loss using \gls{acr:voa}\textsubscript{CH} in \SI{0.5}{dB} steps while the configuration of VOA\textsubscript{TX} and VOA\textsubscript{RX} remains unchanged.
For each attenuation setting of VOA\textsubscript{CH}, we record approximately \num{50} protocol runs.
Sweeping the channel loss from \SIrange{-11.5}{0}{dB}, this totals to around \num{1200} recorded protocol runs.
After removing the runs in which the \gls{acr:dsp} failed to correctly estimate the I/Q skew or in which the received \gls{acr:snr} falls outside the expected range, we are left with \num{864} protocol runs.
To facilitate a fair comparison using the same protocol runs, we introduce the concept of \textit{emulated} transmission distances, i.e., the fiber distance to which the current channel attenuation corresponds to.
Given that the signal powers employed in \gls{acr:cv-qkd}, including the two helper signals, are well below the regime in which effects from fiber non-linearity become relevant, the quantum channel can be modeled as a \textit{block-wise} \gls{acr:lti} system.
The effects from long-term component drifts, such as laser power or \gls{acr:voa} attenuation, are accounted for through the periodic calibration of the experimental setup at the beginning of each protocol run.
Short-term time-varying effects from components can be partially compensated by \gls{acr:dsp}.
For instance, polarization rotations are monitored and mitigated by an adaptive equalizer while clock deviations between \acrshort{acr:awg} and \acrshort{acr:dso} are resolved using interpolation.
This simplification permits re-distributing the transmittance of VOA\textsubscript{CH} between the three employed \glspl{acr:voa} \textit{after} the measurement run.
While Bob's received waveform remains unchanged, the computation of the Holevo bound and, consequently, the SKR are affected.
Under the trusted-device assumption, part of the channel loss is then attributed to the trusted detector or to a change in modulation variance rather than to a potential attacker.

\begin{figure}
    \centering
    \includegraphics{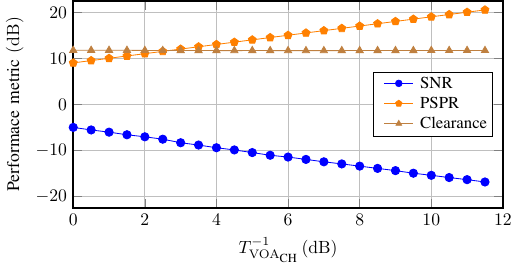}
    \caption{%
        Experimental results for mean \gls{acr:snr}, the configured \gls{acr:pspr} and mean clearance as a function of \gls{acr:voa}\textsubscript{CH} attenuation.
    }
    \label{fig:snr_pspr_distance}
\end{figure}

To investigate methods~\circled{1} and \circled{2}, we first group the recorded measurements according to their \gls{acr:voa}\textsubscript{CH} setting.
After removing failed \gls{acr:dsp} runs and considering only protocol runs with positive excess noise, the received \gls{acr:snr} for the considered group of measurements is approximately constant.
In the next step, we re-distribute the constant loss of \gls{acr:voa}\textsubscript{CH} to either \gls{acr:voa}\textsubscript{TX} or \gls{acr:voa}\textsubscript{RX} to emulate changes in distance.
To that end, we compute a vector of emulated channel transmittance, taking into account distances $d$ from \SIrange{0}{100}{km} and assuming a fiber attenuation of \mbox{$\alpha = \SI{0.2}{dB/km}$}.
Given the measured channel transmittance for a particular setting of \gls{acr:voa}\textsubscript{CH}, we determine the baseline, i.e., the true channel attenuation of the current experimental setup.
This is important since the insertion loss of \gls{acr:voa}\textsubscript{CH} and the true attenuation of the \SI{20}{km} \gls{acr:ssmf} fiber spool must be considered.
For method~\circled{1}, the baseline modulation variance is \mbox{$V_{\mathrm{mod}} = \SI{5}{SNU}$}.
If more or less distance should be emulated for a given \gls{acr:voa}\textsubscript{CH} setting, then the emulated modulation variance needs to be adjusted according to
\begin{equation}\label{eq:vmod_emul}
    V_{\mathrm{mod}}^{\mathrm{emul}} = V_{\mathrm{mod}} T_{\text{VOA}\textsubscript{CH}} / T_{\mathrm{FL}} \text{,}
\end{equation}
where $T_{\text{VOA}\textsubscript{CH}}$ denotes the transmittance of \gls{acr:voa}\textsubscript{CH} and \mbox{$T_{\mathrm{FL}} = 10^{-\alpha d}$} is the emulated fiber loss.
Similarly, the emulated attenuation can be re-distributed to \gls{acr:voa}\textsubscript{RX} and be taken into account as trusted detector loss during \gls{acr:skr} computation.
For method~\circled{2}, the detector transmittance is updated as
\begin{equation}\label{eq:tdet_emul}
    T_{\mathrm{det}}^{\mathrm{emul}} = T_{\mathrm{det}} T_{\text{VOA}\textsubscript{CH}} / T_{\mathrm{FL}} \text{,}
\end{equation}
where \mbox{$T_{\mathrm{det}}^{\mathrm{emul}} \in [0, T_{\mathrm{det}}]$}.
Notably, the modulation variance can be chosen freely (within experimentally viable ranges) in method~\circled{1} while the actual trusted detector loss serves as an upper bound on the emulated loss in method~\circled{2}.
At the same time, this is also an upper bound on the achievable (emulated) transmission distance.

\subsection{Experimental Findings}
\begin{figure}
    \centering
    \includegraphics{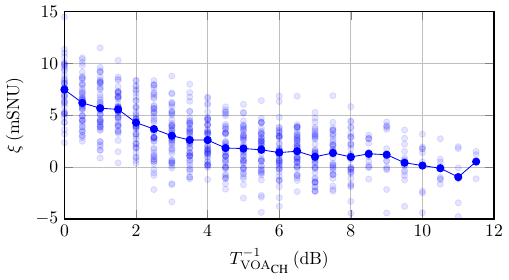}
    \caption{%
        Experimental results for excess noise $\xi$ over the \gls{acr:voa}\textsubscript{CH} attenuation, considering protocol runs with non-failing \gls{acr:dsp} only.
        The means for each \gls{acr:voa}\textsubscript{CH} attenuation setting are indicated with solid markers.
        We report excess noise at the channel output.
    }
    \label{fig:en_distance}
\end{figure}

As shown in \cref{fig:snr_pspr_distance}, the experimental setup operates in the shot-noise limited regime, exhibiting an average clearance of \SI{11.7}{dB} across all protocol runs.
Following the convention in the literature, we compute the clearance as the ratio of shot noise to electronic noise variance, evaluated over the \gls{acr:qkd} signal bandwidth~\cite{Laudenbach2018}.
Moreover, we find that the received \gls{acr:snr} follows the \gls{acr:voa}\textsubscript{CH} attenuation linearly.
The consistent behavior of \gls{acr:snr} and clearance across different VOA\textsubscript{CH} configurations empirically supports the block-wise \gls{acr:lti} assumption, indicating that time-varying effects from components are negligible.
To help the receiver \gls{acr:dsp}, we keep the pilot power at the input of the receiver constant by scaling the \gls{acr:pspr} with the channel loss.
In \cref{fig:en_distance}, we present the measured excess noise~$\xi$ over the \gls{acr:voa}\textsubscript{CH} attenuation, which shows the expected exponential decay over the channel transmittance.
It includes, for instance, noise due to non-ideal state preparation at the transmitter, such as thermal noise in the transmitter electronics.
At the discrete receiver, fiber length mismatches in the eight arms of the polarization-diverse optical hybrid constitute an additional impairment that cannot be mitigated by DSP and that adds to the measured excess noise.
Moreover, DSP implementation imperfections, e.g., an inaccurate estimation and compensation of carrier frequency or clock offset, further contribute to the measured excess noise.
Because of statistical fluctuations, some protocol runs yield negative estimates of the excess noise.
Additionally, the total number of valid protocol runs decreases as channel attenuation increases, since the \gls{acr:dsp} fails more frequently under the more challenging conditions.

\subsection{Information Reconciliation}
We employ an open-source implementation of multidimensional reconciliation~\cite{Leverrier2008} based on a rate-adaptive \gls{acr:rl-ldpc} code~\cite{Cil24,Cil24b}.
Its rate can be adjusted between \num{0.01} and \num{0.2}, and the block length ranges from \SI{1e5}{bit} to \SI{2e6}{bit}.
Eight-dimensional reconciliation is adopted, as it offers the best performance~\cite{Kadir2025} while maintaining low computational complexity due to the Cayley-Dickson construction~\cite{Dickson1919}.
Decoding is performed using the iterative \gls{acr:spa}, with the maximum number of iterations set to \num{150}.
To synthetically increase the number of available states for a protocol run, we iteratively permute and shuffle the recorded quantum states until \num{100} erroneous frames are observed for each protocol run.
This approach is necessary because the number of \gls{acr:fec} frames generated from a single protocol run is too small to yield statistically significant results. 
In addition, especially for low code rates, the block length may even exceed the number of quantum states that are recorded in a single protocol run.

\subsection{Post-processing}
\begin{figure}
    \centering
    \includegraphics{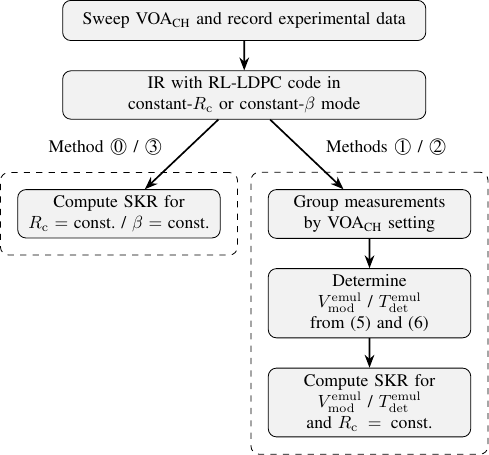}
    \caption{%
        Post-processing of experimental data for methods \protect\circled{0} to \protect\circled{3}.
    }
    \label{fig:postprocessing}
\end{figure}

As the considered experimental setup is modeled as a block-wise \gls{acr:lti} system, re-distributing the attenuation between the employed \glspl{acr:voa} has no impact on Bob's received waveform and on the \gls{acr:ir} phase in general.
We conduct \gls{acr:ir} for all protocol runs with four constant code rates and four different settings for the reconciliation efficiency using the \gls{acr:rl-ldpc} code, either in constant-rate or constant reconciliation efficiency mode.
For the latter, we compute the code rate with the help of~\eqref{eq:mutual_information} and the \gls{acr:snr} estimated after \gls{acr:dsp}.
With the obtained statistics, we determine the extracted \gls{acr:skr} as a function of the emulated distance.
Since higher-order \gls{acr:ps}-QAM has been reported to closely resemble Gaussian modulation~\cite{Denys2021}, we base our analysis on the security proofs for collective attacks under the trusted-device scenario described in~\cite{Laudenbach2019}.
The parameters required to evaluate the Holevo bound, computed according to~\cite{Laudenbach2019}, are determined during the parameter-estimation phase.
The post-processing steps for methods \circled{0} to \circled{3} are summarized in~\cref{fig:postprocessing}.

\section{Results}\label{sec:results}
\begin{figure*}[t]
    \centering
    \begin{minipage}[t]{0.48\textwidth}
        \captionsetup{width=.5\linewidth}
        \centering
        \adjustbox{max width=\textwidth}{
            \includegraphics{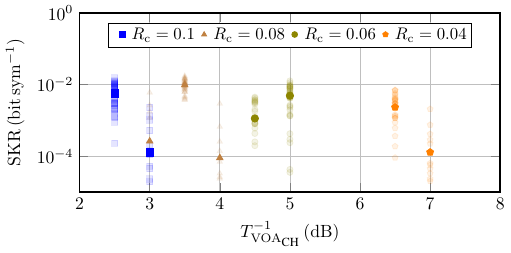}
        }
        \caption{%
            Experimental results for method \protect\circled{0} over the \gls{acr:voa}\textsubscript{CH} attenuation, considering only runs with non-negative excess noise.
            The means for each \gls{acr:voa}\textsubscript{CH} attenuation setting are indicated with solid markers.
        }
        \label{fig:skr-const-rate}
    \end{minipage}%
    \hfill
    \begin{minipage}[t]{0.48\textwidth}
        \captionsetup{width=.5\linewidth}
        \centering
        \adjustbox{max width=\textwidth}{
            \includegraphics{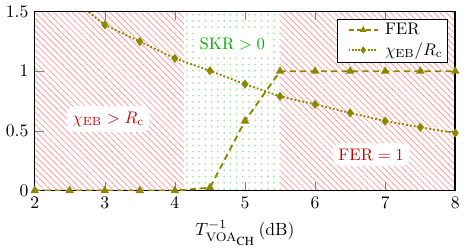}
        }
        \caption{%
            Experimental results for method \protect\circled{0} over the \gls{acr:voa}\textsubscript{CH} attenuation.
            We depict mean \gls{acr:fer} and mean fraction $\chi_{\mathrm{EB}}/R_{\mathrm{c}}$ using the \mbox{$R_{\mathrm{c}} = 0.06$} constant-rate \gls{acr:fec} code for \gls{acr:ir}.
        }
        \label{fig:skr-const-rate-explanation}
    \end{minipage}
\end{figure*}
\begin{figure*}[b]
    \centering
    \begin{minipage}[t]{0.48\textwidth}
        \centering
        \adjustbox{max width=\textwidth}{
            \includegraphics{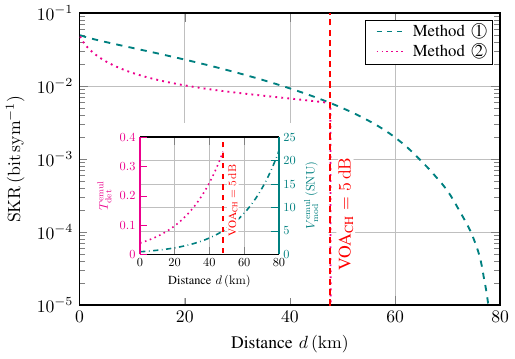}
        }
        \caption{%
            Experimental results for methods \protect\circled{1} and \protect\circled{2} using an \gls{acr:fec} code with rate \mbox{$R_{\mathrm{c}} = 0.06$} over the emulated fiber distance $d$.
            We take into account only experimental runs with non-negative excess noise, and compute the mean extracted \gls{acr:skr} over the group of measurements with the same \gls{acr:voa}\textsubscript{CH} setting.
            The inset plot illustrates the tuned values of $V_{\mathrm{mod}}$ and $T_{\mathrm{det}}$.
        }
        \label{fig:const-rate-opt-vmod-tdet}
    \end{minipage}%
    \hfill
    \begin{minipage}[t]{0.48\textwidth}
        \captionsetup{width=.5\linewidth}
        \centering
        \adjustbox{max width=\textwidth}{
            \includegraphics{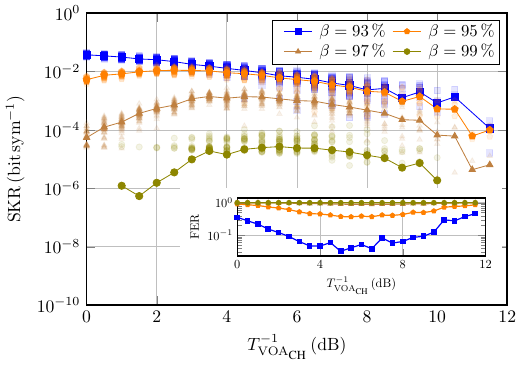}
        }
        \caption{%
            Experimental results for method \protect\circled{3} considering different reconciliation efficiencies over the \gls{acr:voa}\textsubscript{CH} attenuation.
            Only protocol runs with positive excess noise are used for the computation of the extracted \gls{acr:skr}.
            The mean extracted \gls{acr:skr} for each attenuation setting is indicated with a solid marker.
            The inset plot depicts the average \gls{acr:fer}.
        }
        \label{fig:skr-adpt-rate}
    \end{minipage}
\end{figure*}

To experimentally validate the findings on distance limitations associated with constant-rate \gls{acr:fec}, we perform \gls{acr:ir} across a range of \gls{acr:voa}\textsubscript{CH} settings.
The results for code rates \mbox{$R_{\mathrm{c}} = \{0.04, 0.06, 0.08, 0.1\}$}, shown in \cref{fig:skr-const-rate}, demonstrate that a positive extracted \gls{acr:skr} is only achievable within distinct attenuation ranges or, equivalently, distance windows when converted to \gls{acr:ssmf} length.
As exemplified for the \mbox{$R_{\mathrm{c}} = 0.06$} \gls{acr:fec} code in \cref{fig:skr-const-rate-explanation}, the \gls{acr:fer} rises significantly, approaching one, for attenuations larger than \SI{5}{dB}.
Conversely, for an attenuation less than \SI{4.5}{dB}, the ratio $\chi_{\mathrm{EB}}/R_{\mathrm{c}}$ increases and eventually exceeds one, implying that Eve gains more information on the shared secret than Alice and Bob.
Thus, secret key extraction using the \mbox{$R_{\mathrm{c}} = 0.06$} constant-rate \gls{acr:fec} code is feasible only within the narrow attenuation range around \SI{4.5}{dB} to \SI{5}{dB}.
Using a set of constant-rate \gls{acr:fec} codes as in method~\circled{0}, the \gls{acr:voa}\textsubscript{CH} attenuation ranges from approximately \SI{2.5}{dB} to \SI{4}{dB}, \SI{4.5}{dB} to \SI{5}{dB} and \SI{6.5}{dB} to \SI{7}{dB} can be covered with the investigated codes.

Although methods~\circled{1} and \circled{2}, i.e., tuning modulation variance and adding controlled amounts of trusted detector loss, still rely on a single \gls{acr:fec} code with fixed rate, they extend the operational range of the system.
\Cref{fig:const-rate-opt-vmod-tdet} shows the extracted \gls{acr:skr} for a constant code rate of \mbox{$R_{\mathrm{c}} = 0.06$}, tuning modulation variance and trusted detector loss for distance-adaptivity.
As illustrated, both approaches enable secure key extraction up to approximately \SI{47.7}{km} of emulated fiber distance, corresponding to the \gls{acr:voa}\textsubscript{CH} setting of the considered experimental runs.
Beyond this point, further adjustment of the trusted detector loss is no longer feasible, as it would imply an unphysical improvement of the detector efficiency.
Under the given experimental conditions, continued scaling of the modulation variance enables the system to sustain a positive \gls{acr:skr} up to \SI{79}{km} of emulated fiber distance.
Throughout tuning both parameters, their values remain within experimentally viable ranges, as shown in the inset of \cref{fig:const-rate-opt-vmod-tdet}.

For method~\circled{3}, we investigate the use of rate-adaptive error-correcting codes.
The extracted \gls{acr:skr} is shown in \cref{fig:skr-adpt-rate} for reconciliation efficiencies \mbox{$\beta = \{0.93, 0.95, 0.97, 0.99\}$}.
As depicted, the rate-adaptive \gls{acr:fec} codes with \SI{93}{\percent}, \SI{95}{\percent} and \SI{97}{\percent} reconciliation efficiency yield positive extracted \glspl{acr:skr} across the entire range of considered attenuation settings.
In contrast, the rate-adaptive \gls{acr:fec} code with \SI{99}{\percent} reconciliation efficiency only allows for secret key extraction when the attenuation is larger than \SI{0.5}{dB} and smaller than \SI{10.5}{dB}.
Although this configuration achieves the highest asymptotic \gls{acr:skr}, i.e., \mbox{$\beta I _{\mathrm{AB}} - \chi_{\mathrm{EB}}$}, the extracted \gls{acr:skr} experiences a severe penalty due to the high \gls{acr:fer} as illustrated in the inset of \cref{fig:skr-adpt-rate}.
Contrarily, the \gls{acr:fer} of the \gls{acr:fec} code with \mbox{$\beta = 0.93$} remains below an \gls{acr:fer} of \num{0.5} throughout all evaluated attenuation settings.
Therefore, despite having the lowest asymptotic \gls{acr:skr} among the examined rate-adaptive \gls{acr:fec} codes, its reduced \gls{acr:fer} penalty outweighs the smaller asymptotic key rate, yielding the highest extracted \gls{acr:skr} of all considered rate-adaptive \gls{acr:fec} schemes.
To maximize reach, the reconciliation efficiency should be chosen as large as possible, given that the \gls{acr:fer} is smaller than  one, since it ultimately defines when the difference $\beta I_{\mathrm{AB}} - \chi_{\mathrm{EB}}$ becomes zero.

\begin{figure*}
    \centering
    \adjustbox{max width=\textwidth}{
        \includegraphics{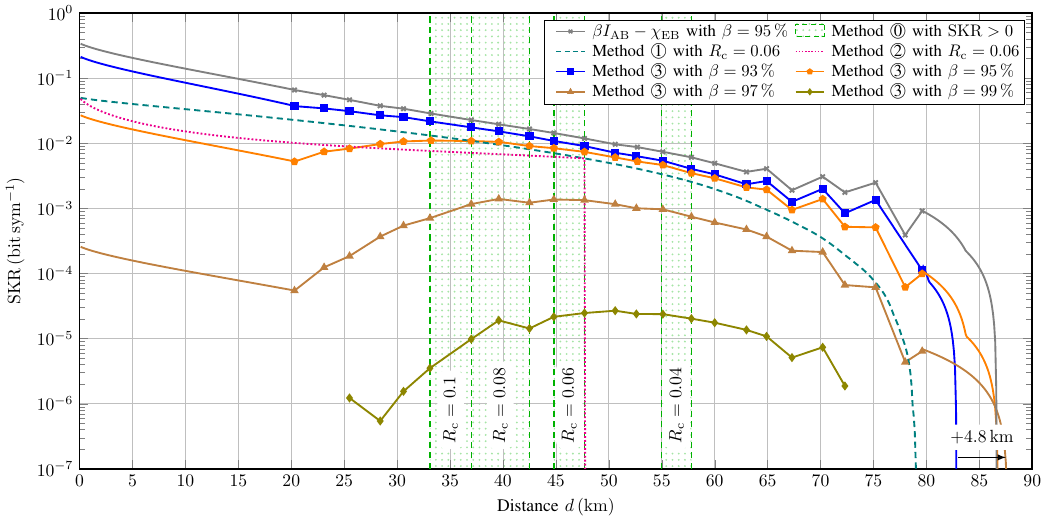}
    }
    \caption{%
        Experimental results for mean extracted secret key rate over the emulated fiber distance $d$.
        We depict a reference curve, i.e., \mbox{$\beta I_{\mathrm{AB}} - \chi_{\mathrm{EB}}$} with \mbox{$\beta = \SI{95}{\percent}$}, and the four investigated approaches to distance-adaptivity.
        For all methods, we compute the extracted \gls{acr:skr} only for experimental runs with non-negative excess noise.
    }
    \label{fig:skr_distance}
\end{figure*}

In \cref{fig:skr_distance}, we present the performance of all investigated methods over emulated fiber distance.
For the sake of a fair comparison and due to experimental limitations, we adjust the modulation variance of approaches using rate-adaptive \gls{acr:fec} for distances below \SI{20}{km} and beyond \SI{80}{km} similar to approach \circled{1} by scaling the modulation variance according to \eqref{eq:vmod_emul}.
To determine the extracted \gls{acr:skr}, we use the closest (in terms of distance) experimental results, i.e., the protocol runs at approximately \SI{20}{km} and \SI{80}{km}.
As reference, we depict the asymptotic \gls{acr:skr}, i.e., \mbox{$\beta I_{\mathrm{AB}} - \chi_{\mathrm{EB}}$} for a fixed reconciliation efficiency of \SI{95}{\percent}.
Effectively, this represents an ideal rate-adaptive error-correcting code with zero residual frame errors and distance-independent reconciliation efficiency.

Due to the discrete attenuation steps of the considered experimental setup, we depict the green-shaded distance regions for the \gls{acr:fec} codes of method~\circled{0} where positive extracted \glspl{acr:skr} are obtained.
As shown, the chosen constant-rate \gls{acr:fec} codes are able to cover distance windows up to a maximum width of \SI{5.5}{km}.
Therefore, method~\circled{0}, i.e., employing all four constant-rate \gls{acr:fec} codes, is able to extract secrecy only for a small subset of considered distances.
Covering the full distance range requires a considerably higher number of \gls{acr:fec} codes since the distance windows become narrower for higher transmission distances.

When comparing methods~\circled{1} and \circled{2}, both of which employ a single constant-rate \gls{acr:fec} code, a performance penalty relative to the reference curve is observed.
For method~\circled{1}, the penalty is smaller than for method~\circled{2}, reaching around one order of magnitude in \gls{acr:skr} for the latter below \SI{10}{km}.
In addition, the secret key extraction of method~\circled{2} is limited to distances smaller than \SI{47.7}{km}, whereas tuning the modulation variance in method~\circled{1} facilitates continuous operation up to \SI{79}{km}.
Since we keep the received \gls{acr:snr} constant over distance by adjusting either the modulation variance or the trusted detector loss and the same \gls{acr:fec} code is used, both methods achieve a distance-independent reconciliation efficiency of \SI{95.96}{\percent}.
Therefore, at an emulated distance of \SI{47.7}{km}, where the modulation variance and the trusted detector loss are the same for methods~\circled{1} to \circled{3}, the extracted \gls{acr:skr} lies in between those of the rate-adaptive method with \mbox{$\beta = \SI{95}{\percent}$} and \mbox{$\beta = \SI{97}{\percent}$}.

Method~\circled{3}, which employs rate-adaptive \gls{acr:fec}, operates close to the reference curve for distances between \SI{40}{km} and \SI{70}{km}, where the \glspl{acr:fer} of the rate-adaptive \gls{acr:fec} codes with reconciliation efficiencies of \SI{93}{\percent} and \SI{95}{\percent} are relatively small.
Outside this range, the performance penalty increases, reaching roughly one order of magnitude for distances below \SI{20}{km} when considering the \gls{acr:fec} code with \mbox{$\beta = 0.95$}.
In contrast, due to the significantly lower \gls{acr:fer} exhibited by the \gls{acr:fec} code with \mbox{$\beta = 0.93$}, the gap between the (zero-\gls{acr:fer}) reference curve and the extracted \gls{acr:skr} is substantially smaller.
Using rate-adaptive \gls{acr:fec} with higher reconciliation efficiencies allows for longer reach, but also introduces larger performance penalties for shorter distances.
For instance, the rate-adaptive \gls{acr:fec} code with \SI{97}{\percent} reconciliation efficiency achieves an additional \SI{4.8}{km} transmission distance compared to the code with \SI{93}{\percent} reconciliation efficiency.
However, this improvement comes at the expense of an \gls{acr:skr} penalty of three orders of magnitude below \SI{20}{km} of emulated fiber distance.
Due to the high \gls{acr:fer} of the rate-adaptive \gls{acr:fec} code with \mbox{$\beta = 0.99$}, it is only able to extract key material from \SI{25.5}{km} to \SI{72.3}{km}, being the narrowest distance window and featuring the lowest extracted \gls{acr:skr} of methods~\circled{1} to \circled{3}.

\section{Discussion}\label{sec:discussion}
The inherent distance limitations associated with constant-rate \gls{acr:fec} codes restrict the use of \gls{acr:cv-qkd} systems to small windows of consecutive distances. 
Using constant-rate \gls{acr:fec} with a \textit{single} code is viable only if the operating distance of the system is known in advance and falls within an operational distance window. 
As an alternative approach, a potentially extensive set of constant-rate \gls{acr:fec} codes, each tailored to cover a specific target distance window, could be used.
However, implementing this approach necessitates numerous code designs and entails more hardware utilization since the parity-check matrices of all \gls{acr:fec} codes need to be stored.
In addition, even small variations in the received \gls{acr:snr} during runtime can have a significant impact on the extracted \gls{acr:skr}, especially when the system operates near the zero-key threshold.
Therefore, using constant-rate \gls{acr:fec} for \gls{acr:ir} is suboptimal without further adjustments.

Tuning the modulation variance or adding controlled amounts of trusted detector loss can enable distance-adaptive operation using a single constant-rate \gls{acr:fec} code.
With respect to implementation, methods~\circled{1} and \circled{2} can be realized based on the information locally available to Alice and Bob.
For method~\circled{1}, the received \gls{acr:snr} is known to Alice through \gls{acr:ir}.
With this knowledge, Alice is able to adjust the modulation variance using \gls{acr:voa}\textsubscript{TX}.
In method~\circled{2}, Bob has access to the received \gls{acr:snr} after parameter estimation and can therefore apply the correct amount of trusted detector loss by adjusting the attenuation of \gls{acr:voa}\textsubscript{RX}.
Since both \glspl{acr:voa} are essential to implement the \gls{acr:cv-qkd} protocol, i.e., either by attenuating the \gls{acr:qkd} signal to the quantum level or by implementing the shutter during calibration, there is no additional cost on hardware.
If the accompanying \gls{acr:skr} penalty is tolerable, these methods increase the operational range considerably and are a simple method to overcome the distance limitations otherwise imposed by constant-rate \gls{acr:fec}.
Due to the practical limitations in adjusting the trusted detector loss, tuning the modulation variance is better suited for long-distance \gls{acr:cv-qkd}.

When the reconciliation efficiency is selected appropriately, the rate-adaptive \gls{acr:fec} utilized in method~\circled{3} achieves \glspl{acr:skr} that approach the asymptotic reference curve.
Unlike methods~\circled{1} and \circled{2}, in which Alice or Bob need to modify parameters of the quantum transmission phase, method~\circled{3} requires no adjustments on the physical layer, thereby enabling joint optimization of protocol parameters and \gls{acr:fec}.
This benefit comes at the expense of additional complexity in the decoder implementation.
Although a single decoder architecture is sufficient for \gls{acr:rl-ldpc} codes, selecting the appropriate subgraph corresponding to the given code rate increases the design complexity.
Furthermore, the hierarchical decoder structure constrains the block lengths as lower-rate codes necessitate larger block lengths that cannot be independently configured.
Therefore, choosing a specific code rate and block length also determines the block lengths of adjacent rates.
As comparable requirements for rate-adaptive \gls{acr:fec} exist in mobile networks, this class of decoders has undergone standardization for 5G NR~\cite{Richardson2018}.

To ensure a fair performance comparison of the analyzed methods, we performed \gls{acr:ir} on recorded measurement data.
In a practical system, however, methods~\circled{1} and \circled{2} can only be applied prior to the subsequent protocol run, not retroactively after a run has completed.
Therefore, the temporal stability of the investigated \gls{acr:cv-qkd} system and its sensitivity to parameter changes are of particular interest to ensure that the adjusted parameters actually yield non-zero \glspl{acr:skr} in the next protocol run.
One possible realization to mitigate these shortcomings is to characterize the \gls{acr:cv-qkd} system using an analytical model, and to implement a feedback loop between Alice and Bob to optimize the modulation variance as demonstrated in~\cite{Ma2023}.
We expect that a similar method could also be applied to method~\circled{2}.
In contrast, since method~\circled{3} requires no changes on the physical layer, its application to a practical system is simpler as the code rate (or equivalently the reconciliation efficiency) can be chosen independently after the quantum transmission phase.
It would therefore be better suited for free-space applications, as retrospective tuning of system parameters is not feasible due to rapidly varying channel conditions~\cite{Kadir2025}.

\section{Conclusion}\label{sec:conclusion}
In this work, we analyze the fundamental limitations of \gls{acr:cv-qkd} employing constant-rate \gls{acr:fec} codes.
Our analysis shows that a positive \gls{acr:skr} can only be achieved within constraint distance windows where two conditions are fulfilled: (i) the \gls{acr:fer} is less than one, and (ii) the code rate is larger than the Holevo bound.
By employing a set of constant-rate \gls{acr:fec} codes, distance-adaptive operation can be realized, albeit at the cost of requiring numerous such codes.
To overcome this limitation, we investigate three alternative strategies for distance-adaptive \gls{acr:cv-qkd}.
The first two methods, i.e., tuning modulation variance or adding controlled amounts of trusted detector loss, can facilitate continuous operation without restrictive distance windows using a single constant-rate \gls{acr:fec} code.
With the investigated experimental setup, secret key extraction is possible up to \SI{47.7}{km} when controlled amounts of trusted detector loss are added, or up to \SI{79}{km} when the modulation variance is tuned.
As an alternative, we study rate-adaptive \gls{acr:fec} and find that operation near the asymptotic \gls{acr:skr} can be achieved over a wide range of distances, provided that the trade-off between \gls{acr:fer} and reconciliation efficiency is selected appropriately.
In future work, joint optimization of system parameters, such as modulation variance and trusted detector loss, in combination with rate-adaptive \gls{acr:fec} could be explored experimentally to maximize the \gls{acr:skr} and extend the reach of \gls{acr:cv-qkd} systems.

\printbibliography

\end{document}